\documentclass[conference,anonymous]{IEEEtran}
\usepackage[utf8]{inputenc}
\usepackage[english]{babel}
\PassOptionsToPackage{bookmarks=false}{hyperref}
\usepackage{multirow, multicol, booktabs, tabulary, tabu, longtable, array, varwidth}
\usepackage{placeins, lipsum}
\usepackage[flushleft]{threeparttable}

\usepackage[hyphens]{url}
\usepackage[labelfont=bf]{caption}
\usepackage{cite}
\usepackage{hyperref}
\usepackage[dvipsnames, table]{xcolor}
\hypersetup{
	linkcolor  = violet!85!black,
	citecolor  = magenta!85!black,
	urlcolor   = blue!85!black,
	colorlinks = true,
	breaklinks = true
}

\usepackage{mathrsfs}

\usepackage{enumerate}

\usepackage[inline]{enumitem}

\usepackage{amsmath, amsfonts, amssymb, amsthm, nicefrac}
\usepackage{siunitx}
\theoremstyle{definition}
\newtheorem{definition}{Definition}

\usepackage{listings}

\usepackage[scaled=0.82]{beramono}
\usepackage{xspace}

\usepackage[]{cleveref} \crefname{appsec}{Appendix}{Appendices}
\crefformat{section}{\S#2#1#3}
\crefformat{subsection}{\S#2#1#3}
\crefformat{subsubsection}{\S#2#1#3}
\crefformat{equation}{(#2#1#3)}
\crefrangeformat{equation}{(#3#1#4--#5#2#6)}
\crefmultiformat{equation}{(#2#1#3)}{ and~(#2#1#3)}{, (#2#1#3)}{ and~(#2#1#3)}
\crefformat{figure}{Fig.~#2#1#3}
\crefrangeformat{figure}{Figs. #3#1#4--#5#2#6}
\crefmultiformat{figure}{Figs.~#2#1#3}{ and~#2#1#3}{, #2#1#3}{ and~#2#1#3}
\crefname{algocf}{Alg.}{Algs.}
\Crefname{algocf}{Algorithm}{Algorithms}
\crefformat{table}{Table~#2#1#3}
\crefrangeformat{table}{Tables~#3#1#4--#5#2#6}
\crefmultiformat{table}{Tables~#2#1#3}{ and~#2#1#3}{, #2#1#3}{ and~#2#1#3}

\usepackage{textcomp}
\usepackage[shortcuts,acronym]{glossaries}
\usepackage{soul, footnote, xargs}
\usepackage{balance}

\usepackage{graphicx}
\usepackage{subcaption}
\usepackage{tikz, epstopdf, stfloats, bbding, capt-of}
\usepackage{algorithm}
\usepackage{algpseudocode}
\usepackage{microtype}
\usepackage[T1]{fontenc}
\def\baselinestretch{0.95}

\usepackage{balance}
\usepackage{enumitem}
\usepackage{fancyhdr}

\makeatother
\usepackage[utf8]{inputenc}

\newcommand*\circled[1]{\tikz[baseline=(char.base)]{
		\node[shape=circle,draw,inner sep=1.5pt, Maroon, fill=Maroon] (char)
		{\color{white}\scriptsize\textbf{#1}};}}

\newcommand{\sysname}{RoboTack\xspace}

\newcommand{\dstop}{d_{\text{stop}}}
\newcommand{\dsafe}{d_{\text{safe}}}

 \def\BibTeX{{\rm B\kern-.05em{\sc i\kern-.025em 
b}\kern-.08em
		T\kern-.1667em\lower.7ex\hbox{E}\kern-.125emX}}
\begin{document}
	\makeatletter
	\let\old@ps@headings\ps@headings
	\let\old@ps@IEEEtitlepagestyle\ps@IEEEtitlepagestyle
	\def\confheader#1{%
		\def\ps@headings{%
			\old@ps@headings%
			\def\@oddhead{\strut\hfill#1\hfill\strut}%
			\def\@evenhead{\strut\hfill#1\hfill\strut}%
		}%
		\def\ps@IEEEtitlepagestyle{%
			\old@ps@IEEEtitlepagestyle%
			\def\@oddhead{\strut\hfill#1\hfill\strut}%
			\def\@evenhead{\strut\hfill#1\hfill\strut}%
		}%
		\ps@headings%
	}
	\makeatother
	
	\confheader{%
		Accepted for the 2020 50th Annual IEEE/IFIP International Conference on 
		Dependable Systems and Networks (DSN)
	}
	\title{ML-driven Malware that Targets AV Safety}
	\author{\textit{Anonymized for submission}}
	\author{\IEEEauthorblockN{
			{Saurabh Jha}\IEEEauthorrefmark{1},
			{Shengkun Cui}\IEEEauthorrefmark{1},
			{Subho S. Banerjee}\IEEEauthorrefmark{1},
			{James Cyriac}\IEEEauthorrefmark{1},\\
			{Timothy Tsai}\IEEEauthorrefmark{2},
			{Zbigniew T. Kalbarczyk}\IEEEauthorrefmark{1},
			and
			{Ravishankar K. Iyer}\IEEEauthorrefmark{1}
		}
		\IEEEauthorblockA{\IEEEauthorrefmark{1}{University of Illinois at Urbana-Champaign, 
		Urbana-Champaign, IL 61801, USA.}}
		\IEEEauthorblockA{\IEEEauthorrefmark{2}{NVIDIA Corporation,  Santa Clara, CA 
		94086, USA.}}
	}

	\maketitle
\begin{abstract}
Ensuring the safety of autonomous vehicles (AVs) is critical for their mass
deployment and public adoption. However, security attacks that violate safety
constraints and cause accidents are a significant deterrent to achieving public
trust in AVs, and that hinders a vendor's ability to deploy AVs. Creating a
security hazard that results in a severe safety compromise (for example, an
accident) is compelling from an attacker's perspective. In this paper, we
introduce an attack model, a method to deploy the attack in the form of smart
malware, and an experimental evaluation of its impact on production-grade
autonomous driving software. We find that determining the time interval during
which to launch the attack is{ critically} important for causing safety hazards
(such as collisions) with a high degree of success. For example, the smart
malware caused $\mathbf{33\times}$ more forced emergency braking than random
attacks did, and accidents in 52.6\% of the driving simulations. 
\end{abstract} 
\section{Introduction}\label{s:intro}
Autonomous vehicle (AV) technologies are advertised to be transformative, with
the potential to bring greater convenience, improved productivity, and safer
roads~\cite{gerla2014internet}. Ensuring the safety of AVs is critical for their
mass deployment and public
adoption~\cite{jha2018avfi,banerjee2018hands,
jhakayotee,hutchison2018robustness,
drivefi,koopman2019safety}. However, security attacks that violate safety
constraints and cause accidents are a significant deterrent to achieving public
trust in AVs, and also hinder vendors' ability to deploy AVs. Creating a
security hazard that results in a serious safety compromise (for example, an
accident) is attractive from an attacker's perspective. For example, smart
malware can modify sensor data at an opportune time to interfere with the
inference logic of an AV's perception module. The intention is to miscalculate
the trajectories of other vehicles and pedestrians, leading to unsafe driving
decisions and consequences. Such malware can fool an AV into inferring that an
in-path vehicle is moving out of the lane while in reality the vehicle is
slowing down; that can lead to a serious accident. 

This paper introduces i) the foregoing attack model, ii) a method to deploy the
attack in the form of smart malware (\sysname), and iii) an experimental
evaluation of its impact on production-grade autonomous driving software.
Specifically, the proposed attack model answers the questions of {\it what},
{\it how}, and {\it when} to attack. The key \textit{research questions}
addressed by \sysname and the \textit{main contributions} of this paper are:

\textbf{Deciding what to attack?}
\sysname modifies sensor data of the AV such that the trajectories of other
vehicles and pedestrians will be miscalculated. \sysname leverages situation
awareness to select a target object for which the trajectory will be altered.

\textbf{Deciding how to attack?}
\sysname minimally modifies the pixels of one of the AV's camera sensors using
an adversarial machine-learning (ML) technique (described in \cref{s:adv_patch})
to alter the trajectories of pedestrians and other vehicles, and maintains these
altered trajectories for a short time interval.  The change in the sensor image
and the perceived trajectory is small enough to be considered as noise. Moreover,
\sysname overcomes compensation from other sensors (e.g., LIDAR) and temporal
models (e.g., Kalman filters).

\textbf{Deciding when to attack?}
\sysname employs a shallow 3-hidden-layered neural network (NN) decision model
(described in \cref{s:time_of_attack}) to identify the most opportune time with
the intent of causing a safety hazard (e.g., collisions) with a high probability
of success. In particular, the proposed NN models the non-linear relationship
between the AV kinematics (i.e., distance, velocity, acceleration) and attack
parameters (i.e., when and how long to attack). We use a feed-forward NN because
neural networks can approximate complex continuous functions as shown in the universal
function approximation theorem~\cite{csaji2001approximation}. 

\textbf{Assessment on production software.} We deployed \sysname on
Apollo~\cite{apollo}, a production-grade AV system from Baidu, to quantify the
effectiveness of the proposed safety-hijacking attack by simulating $^\sim 2000$
runs of experiments for five representative driving scenarios using the LGSVL
simulator~\cite{lgsvl}.

The {\it key findings} of this paper are as follows:
\begin{enumerate}[noitemsep,nolistsep,leftmargin=*]
    
    \item \sysname is significantly more successful in creating safety hazards than random attacks (our baseline) are. Here, random attacks correspond to miscalculations of the trajectories (i.e., {\it trajectory hijacking}) of randomly chosen non-AV vehicles or pedestrians, at random times, and for random durations. This random attack condition is the most general condition for comparison, although we also show results for a much more restrictive set of experiments. \sysname caused $\mathbf{33\times}$ more forced emergency braking than random attacks did, i.e., \sysname caused forced emergency braking in \textbf{75.2\%} of the runs (640 out of 851). In comparison, random attacks caused forced emergency braking in \textbf{2.3\%} (3 out of 131 driving simulations).\footnote{These numbers, while seemingly contradictory, are consistent, as we will show in \cref{s:results}.} 
    
    \item Random attacks caused \textbf{0} accidents, whereas \sysname caused
    accidents in \textbf{52.6\%} of the runs (299 out of 568). 
    
    \item \sysname had higher a success rate in attacking pedestrians
    (\textbf{84.1\%} of the runs that involved pedestrians) than in attacking
    vehicles (\textbf{31.7\%} of the runs that involved vehicles). 
    
    \item Apollo's perception system is less robust in detecting pedestrians
    than in detecting other vehicles.  \sysname \textit{automatically} discerns
    that difference, and hence needs  {\it only} \textbf{14} consecutive camera
    frames involving pedestrians to cause accidents, while needing \textbf{48}
    consecutive camera frames that only involve other vehicles to cause
    accidents.
    
\end{enumerate}

{\bf Comparing \sysname with adversarial learning.} Past work has targeted deep
neural networks (DNNs) used in the perception systems of AVs to create
adversarial machine-learning-based
attacks~\cite{cao2019adversarial,boloor2019attacking,
eykholt2017robust,lu2017adversarial} that were shown to have successful results
(such as by causing misclassification and/or misdetection of a stop sign as a
yield sign), and object trackers~\cite{jia2019fooling}. The goal of this line of
research is to create adversarial objects on the road that fool the AV's
perception system. However, such attacks
\begin{enumerate*}
\item are limited because DNNs represent only a small portion of the production
autonomous driving system (ADS)~\cite{apollo}, and 
\item have low safety impact due to built-in compensation provided by temporal
state-models (which provide redundancy in time) and sensor fusion (which
provides redundancy in space) in ADS, which can mask the consequences of such
perturbations and preserve AV safety (as shown in this paper, and by
others~\cite{lu2017no}).
\end{enumerate*}
To summarize, adversarial learning tells one only {\it what} to attack. In
contrast, as we discuss in detail in \cref{s:attackphases}, \sysname tells one
{\it what, when}, and {\it how} to attack, making it highly efficient in
targeting AV safety. Moreover, previous attacks have been shown {\it only}
on one camera sensor without considering i) the sensor fusion module, and ii)
the AV control loop (i.e., it considers only statically captured video
frames without running a real ADS). In contrast, we show our attack on an end-to-end 
production-grade AV system using a simulator that provides data from multiple sensors. 
 \begin{figure*}[!t]
    \centering
    \includegraphics[width=0.82\textwidth]{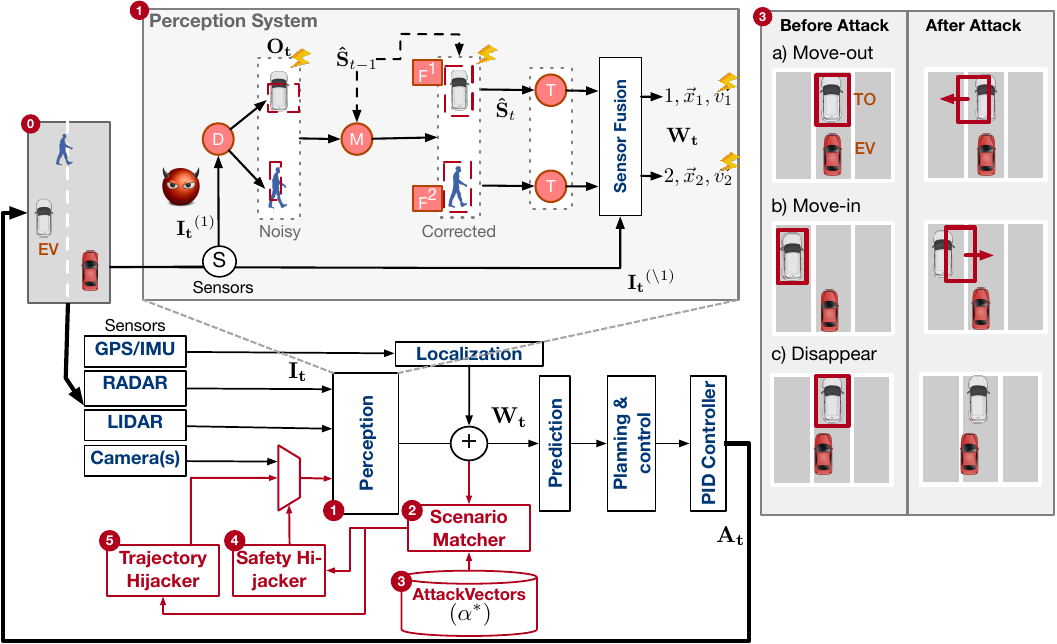}
    \caption{Overview of the ADS perception system and the proposed attack in \sysname.}
    \label{fig:overview_ads_attack}
\end{figure*}

\vspace{-0.1cm}
\section{Background}~\label{s:background}
\subsection{Autonomous Driving Software}
We first discuss the terminologies associated with the autonomous driving system
(ADS) that are used in the remainder of the paper.
\cref{fig:overview_ads_attack} illustrates the basic control architecture of an
AV (henceforth also referred to as the \emph{Ego vehicle}, EV). The EV consists
of mechanical components (e.g., throttle, brake, and steering) and actuators
(e.g., electric motors) that are controlled by an \emph{ADS}, which represent
the computational (hardware and software) components of the EV. At every instant
in time, $t$, the ADS system takes input from sensors $\mathbf{I_t}$ (e.g.,
cameras, LiDAR, GPS, IMU) and infers $\mathbf{W_t}$, a model of the world, which
consists of the positions and velocities of objects around the EV. Using
$\mathbf{W_t}$ and the destination as input, the ADS planning, routing, and
control module generates actuation commands (e.g., throttle, brake, steering
angle). Those commands are smoothed out using a PID
controller~\cite{aastrom1995pid} to generate final actuation values
$\mathbf{A_t}$ for the mechanical components of the EV. The PID controller
ensures that the AV does not make any sudden changes in  $\mathbf{A_t}$.

\subsection{Perception System}
\begin{definition}
     {\it Object tracking} is defined as the process of identifying an object
     (e.g., vehicle, pedestrian) and estimating its state $s_t$ at time $t$
     using a series of sensor measurements (e.g., camera frames, LIDAR
     pointcloud) observed over time. The \emph{state} of the object is
     represented by the coordinates and the size of a ``bounding box''
     (\textit{bbox}) that contains the object. That estimated state at time $t$
     is used to estimate the trajectory (i.e., the velocity, acceleration, and
     heading) for the object.
     \label{def:tracking}
\end{definition}

\begin{definition}
    \emph{Multiple object tracking} (MOT) is defined as the process of estimating the state of the world denoted by $\mathbf{\hat{S}}_t = (\hat{s}_t^1, \hat{s}_t^2, ..., \hat{s}_t^{N_t})$, where $N_t$ represents the number of objects in the world at time $t$, and $\hat{s}_t^i$ is the state of the $i$\textsuperscript{th} object.\footnote{In this paper, the boldface math symbols represent tensors and nonbolded symbols represent scalar values in tensors.}
\end{definition}

The MOT problem is most commonly solved by the {\it tracking-by-detection}
paradigm~\cite{DBLP:journals/corr/LuoZK14}. An overview of this paradigm is
shown in \cref{fig:overview_ads_attack}. Here, a sensor (or group of sensors)
continuously collects the measurement data at every instant of time $t$
($\mathbf{I}_t$). These sensor inputs are sent to a corresponding DNN-based
\emph{object detector}, such as YoloNet~\cite{redmon2018yolov3} or
FasterRCNN~\cite{ren2015faster} (labeled as ``D'' in
\cref{fig:overview_ads_attack}). Such an object detector estimates the object's
class and its bbox at every time instant. The collection of these bbox
measurements for all objects is denoted by $\mathbf{O_t} = \{o^1_t, o^2_t, ...,
o^{M_t}_t\}$, where  $o_t^i$ denotes the observations for the
$i$\textsuperscript{th} object at time $t$.

An \emph{object tracker} (or just \emph{tracker}) tracks the changes in the
position of the bboxes over successive sensor measurements. Each detected object
is associated with a unique tracker, where a tracker is a Kalman
filter~\cite{welch1995introduction} (KF) that maintains the state $s^i$ for the
$i$\textsuperscript{th} object. Each object detected at time $t$ is {\it
associated} with either an existing object tracker or a new object tracker,
initialized for that object. Such association of a detected object with existing
trackers (from time $t-1$) is formulated as a bipartite matching problem, which
is solved using the Hungarian matching algorithm~\cite{huang2008robust} (shown
as ``M'' in the figure). ``M'' uses the overlap, in terms of
IoU\footnote{Intersection over Union (IoU) is a metric for characterizing the
accuracy of predicted bounding boxes (bboxes). It is defined as
$\nicefrac{(\text{area of overlap})}{(\text{area of union})}$ between the
ground-truth label of the bbox and the predicted bbox.}, between the detected
bboxes at time $t$ (i.e., the output of ``D'') and the bboxes predicted by the
trackers (Kalman filter) of the existing objects to find the matching. A KF is
used to maintain the temporal state model of each object (shown as
``F\textsuperscript{*}'' in the figure), which operates in a recursive
predict-update loop: the predict step estimates the current object state
according to a motion model, and the update step takes the detection results
($o^{i}_{t}$) as the measurement to update the $\hat{s}^i_t$ state. That is, the
KF uses a series of noisy measurements observed over time and produces estimates
of an object state that tend to be more accurate than those based on a single
measurement alone. KF is used to address the following challenges:

\begin{itemize}[noitemsep,nolistsep,leftmargin=*]
	\item Sensor inputs are captured at discrete times (i.e., $t,~t+1,\dots$).
	Depending on the speed and acceleration, the object may have moved between
	those discrete time intervals. Motion models associated with KFs predict the
	new state of tracked objects from time-step $t-1$ to $t$.
	\item State-of-the-art object detectors are inherently
	noisy~\cite{redmon2018yolov3,ren2015faster} (i.e., bounding box estimates
	are approximate measurements of the ground truth), and that can corrupt the
	object trajectory estimation (i.e., velocity, acceleration, heading). Hence,
	the perception system uses KFs to compensate for the noise by using Gaussian
	noise models~\cite{huang2008robust}.
\end{itemize} 

Finally, a transformation operation (shown as ``T'' in the figure) estimates
the position, velocity, and acceleration for each detected object by using
$\mathbf{\hat{S}_{t}}$. The transformed measurements are then fused with other
sensor measurements (e.g., from LiDAR) to get world state $\textbf{W}_t$.

\begin{figure}[!t]
	\centering
	\includegraphics[width=0.45\textwidth]{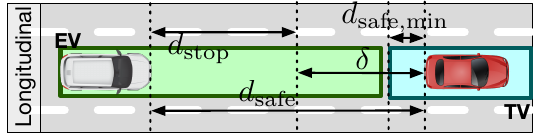}
	\caption{Definition of $\dstop$, $\dsafe$, and $\delta$ for lateral and
	longitudinal movement of the car. Non-AV vehicles are labeled as
	\emph{target vehicles} (TV).}
	\label{fig:safety}
\end{figure}

\subsection{Safety Model}\label{s:safety_model}
In this paper, we use the AV safety model provided by Jha {et
al.}~\cite{drivefi}. They define the instantaneous safety criteria of an AV in
terms of the longitudinal (i.e., direction of the vehicle's motion) and lateral
(i.e., perpendicular to the direction of the vehicle's motion) Cartesian
distance travelled by the AV (see \cref{fig:safety}). In this paper, we use only
the longitudinal definition of the safety model, as our attacks are geared
towards driving scenarios for which that is relevant. Below we reproduce the
definitions of Jha {et al.}'s safety model for completeness.
\begin{definition}
	The \emph{stopping distance} $\dstop$ is defined as the maximum distance the
	vehicle will travel before coming to a complete stop, given the maximum
	comfortable deceleration.
\end{definition}

\begin{definition}
	The \emph{safety envelope} $\dsafe$~\cite{erlien2015shared,suh2016design} of
	an AV is defined as the maximum distance an AV can travel without colliding
	with any static or dynamic object.
\end{definition}

In our safety model, we compute $\dsafe$ whenever an actuation command is sent
to the mechanical components of the vehicle. ADSs generally set a minimum value
of $\dsafe$ (i.e., $d_{\text{safe}, \text{min}}$) to ensure that a human
passenger is never uncomfortable about approaching obstacles.

\begin{definition}
\label{def:potential}
	The \emph{safety potential} $\delta$ is defined as $\delta = \dsafe -
    \dstop$. An AV is defined to be in a \emph{safe state} when $\delta > 0$.
\end{definition}
Unlike the authors of \cite{drivefi} who use $\delta \ge 0$ as the safe operation
state, we choose $\delta \ge 4$ meters because of a limitation in the simulation
environment provided by LGSVL~\cite{lgsvl} for Apollo~\cite{apollo} that halts
simulations for distances closer than $4$ meters. \section{Attack Overview \& Threat Model}\label{s:principles}
This section describes the attacker goals, target system, and defender
capabilities.

\subsection{Attacker Goals} \label{s:goals}
The ultimate goal of the attacker is to hijack object trajectories as perceived
by the AV in order to cause a safety hazard. 

To be successful, the attack must:
	
\begin{itemize}[noitemsep,nolistsep,leftmargin=*]
	\item {\bf Stay stealthy by disguising attacks as noise.}
	To evade detection of his or her malicious intent, an attacker may want to
    disguise malicious actions as events that occur naturally during driving. In
    our attack, we hide the data perturbation initiated by the malware/attacker
    as sensor noise. As we show in \cref{s:yolo_char}, modern object detectors
    naturally misclassify (i.e., identify the object class incorrectly) and
    misdetect (i.e., bounding boxes have zero or $<60\%$ IoU) objects for
    multiple time-steps (discussed in \cref{s:yolo_char}). Taking advantage of
    that small error margin in hiding data perturbations, the attacker initiates
    the attack 1) at the \textit{most opportune time} such that even if the
    malicious activity is detected it is too late for the defender to mitigate
    the attack consequences, and 2) for a \textit{short duration of time} to
    evade detection from the intrusion-detection system (IDS) that monitors for
    spurious activities~\cite{cho2016fingerprinting}.
    
	\item {\bf Situational awareness.}
	Hijacking of the object trajectory in itself is not sufficient to cause
    safety violations or hazardous driving situations. An attacker must be aware
    of the surrounding environment to initiate the attack at the most opportune
    time to cause safety hazards (e.g., collisions).
	
	\item {\bf Attack automation.}
	An attacker can automate the process of monitoring and identifying the
	opportune time for an attack. That way, the adversary only needs to install
	the malware instead of manually executing the attack.

\end{itemize}

\subsection{Threat Model} \label{s:threat}
In this section, we discuss the target system, the existing defenses, and the
attacker's capabilities.

\textbf{Target system.}
The target is the perception system of an AV, specifically the object detection,
tracking, and sensor fusion modules. To compensate for the noise in the outputs
of the object detectors, the AV perception system uses temporal tracking and
sensor fusion (i.e., fusion data from multiple sensors such as LIDAR, RADAR,
and cameras). Temporal tracking and sensor fusion provide an inherent defense
against most if not all existing adversarial attacks on
detectors~\cite{lu2017no}.

The critical vulnerable component of the perception system is a \emph{Kalman
filter (KF)} (see ``F'' in \cref{s:background} and
\cref{fig:overview_ads_attack}). KFs generally assume that measurement noise
follows a zero-mean Gaussian distribution, which is the case for the locations
and sizes of bboxes produced by the object detectors (described later in
\cref{s:yolo_char}). However, that assumption introduces a vulnerability. The KF
becomes ineffective in compensating for the adversarially added noise. We show
in this paper that an attacker can alter the trajectory of a perceived object by
adding noise within one standard deviation of the modeled Gaussian noise.

The challenge in attacking a KF is to maintain a small attack window (i.e., the
number of contiguous time epochs for which the data are perturbed). When an
attacker is injecting a malicious noise pattern, the attack window must be
sufficiently small (1--60 time-steps) such that the defender cannot estimate the
distribution of the injected noise and hence cannot defend the system against the attack. 

\noindent \textbf{What can attackers do?}
In this paper we intentionally and explicitly skirt the problem of defining the
threat model. Instead, we focus on what an attacker could do to an AV if he or
she has access to the ADS source code and live camera feeds.

\textit{Gain knowledge of internal ADS system.} We assume that the attacker has
obtained knowledge of the internal ADS system by analyzing the architecture and
source code of open-source ADSs, e.g., Apollo~\cite{apollo,
koscher2010experimental}. The attacker can also gain access to the source code
through a rogue employee.

\textit{Gain access to and modify live camera feed.} 
Recently, Argus~\cite{hackCamera} showed the steps to hijack a standalone
automotive-grade Ethernet camera and spoof the camera traffic. The attack
follows a ``man-in-the-middle'' (MITM) strategy in which an adversary gains
physical access to the camera sensor data and modifies them (when certain
conditions are met). The hack relied on the fact that the camera traffic is
transmitted using standard (i.e., IEEE 802.1 Audio Video
Bridging~\cite{IEEE802190:online}) but simple protocols, which do not use
encryption because of the size of the data as well as performance and latency
constraints associated with the transmission. As the camera feed is not
encrypted, the attacker can reassemble packages of a video frame and decode the
JFIF (JPEG File Interchange Format) payload into an image. Most importantly,
since there is no hash or digital signature checks on the transmitted images, to
prepare for the attack, the attacker can apply a number of filters to modify the
images in-line without being noticed. The MITM attack works by using an
\emph{Ethernet tap} device to capture UDP packets in the Ethernet/RCA link
between the camera and the ADS software. The Ethernet tap captures images and
provides them as the input for attacker-controlled hardware with purpose-built
accelerators, such as NVIDIA EGX, that are operating outside the domain of the
ADS hardware/software.

\textit{Optionally, compromise ADS software using secret hardware implant.} To
further hide malware and evade detection, an attacker can install backdoors in
the hardware. Injection of malicious code in hardware-software stacks has been
realized in existing hardware backdoors embedded in CPUs, networking routers,
and other consumer devices~\cite{winkelman2019autonomous,
koscher2010experimental}. As an AV is assembled from components supplied by
hundreds of vendors through a sophisticated supply chain, it is reasonable to
argue that individual components such as infotainment systems and other existing
electronic component units (ECUs) could be modified to enable secret
backdoors~\cite{InfotainmentHack,koscher2010experimental}.

\textbf{What things can attackers not do?} In this work, we assume that the CAN
bus transmitting the control command is protected/encrypted. Therefore we cannot
launch a man-in-the-middle attack to perturb the control/actuation commands sent
to the mechanical components of the EV.

\textbf{Defender capabilities.}
Again, we assume that the CAN bus transmitting the controller/actuation commands
is encrypted. That assumption is acceptable because many commercial products
utilize such encryption\cite{pfeiffer2017implementing}. Moreover, there are well
known IDSs for monitoring CAN bus
activity~\cite{cho2016fingerprinting,manandhar2014detection}. Therefore, we do
not leverage CAN bus vulnerabilities as an attack vector; instead, our attack
exploits vulnerabilities on the camera's Ethernet/RCA cable link.

\subsection{Attack Vectors and Injected Attacks}\label{subsec:attack_vector}
We describe a taxonomy of attack vectors (shown in
\cref{fig:overview_ads_attack}) that the attacker can leverage to maximize
impact, such as with an emergency stop or a collision. The attack vectors are as
follows.
\begin{itemize}[noitemsep,nolistsep,leftmargin=*]
	\item [\textbf{a)}] {\bf Move\_Out.} In this attack, the attacker hijacks
    the target object (TO) trajectories to fool the EV into believing that the
    TO is moving out of the EV's lane. A close variant of this attack is fooling
    the EV into believing that the target object is maintaining its lane,
    whereas in reality the target object is moving into the EV's lane. Because
    of this attack, the EV will start to accelerate or maintain its speed,
    causing it to collide with the target object.
    	
	\item [\textbf{b)}] {\bf Move\_In.} In this attack, the attacker hijacks the
    target object (TO) trajectories to fool the EV into believing that the TO is
    moving into the EV's lane. Because of this attack, the EV will initiate
    emergency braking. The emergency braking maneuver is highly uncomfortable
    for the passengers of the EV and may lead to injuries in some cases.
	
	\item [\textbf{c)}] {\bf Disappear.} In this attack, the attacker hijacks
    the target object (TO) trajectories to fool the EV into believing that the
    TO has disappeared. The effects of this attack will be similar to those of
    the {\it Move\_Out} attack model.

\end{itemize}

\begin{figure*}[!t]
    \centering
    \includegraphics[width=\textwidth]{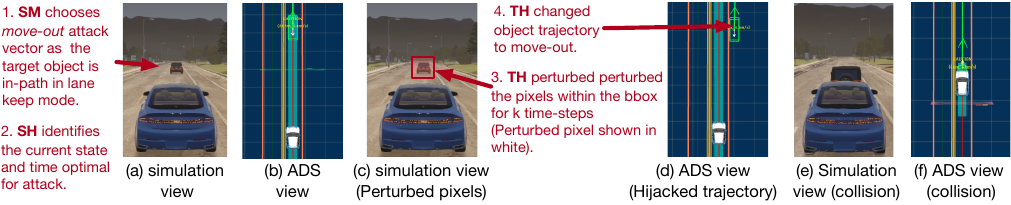}
    \caption{Steps followed by \sysname to mount a successful attack, 
        i.e., collision between the EV (blue) and the target object (red).
     SM: Scenario matching. SH: Safety hijacking, TH: Trajectory hijacking.}
    \label{fig:attack-example}
\end{figure*}
\subsection{Attack Phases.} \label{s:attackphases}

The attack progresses in three key phases as follows.

{\bf Phase 1. Preparing and deploying the malware.} Here the attacker does the
following:
\begin{enumerate}[noitemsep,nolistsep,leftmargin=*]
        \item Gains access to the ADS source code,
        \item Defines the mapping between the attack vectors (see
        \cref{subsec:attack_vector}) and the world state ($\mathbf{W}_t$),
        \item Trains the "safety hijacker" and fine-tunes the weights of
        "trajectory hijacker" (e.g., learns about the maximum perturbation that
        can be injected to evade detection) for the given ADS, 
        \item Gains access to the target EV camera feeds, and 
        \item Installs \sysname on the target EV.
\end{enumerate}

{\bf Phase 2. Monitoring the environment.} Once our intelligent malware is
deployed, it does the following:
\begin{enumerate}[noitemsep,nolistsep,leftmargin=*]
    \item  Approximately reconstructs the world ($\mathbf{{W}}_t$) using the
    hacked camera sensor data (\circled{1} in \cref{fig:overview_ads_attack}).
    For simplicity, we assume that the world state estimated using sensor fusion
    is not significantly different from the state determined using only one
    camera sensor.  
    In our implementation, we only use $\mathbf{\hat{S}}_t$ to carry out the attack
    instead of relying on data from all sensors.
    
    \item Identifies the victim target object $i$ (i.e., one of the other
    vehicles or pedestrians) for which the trajectory is hijacked. The target
    object is the one closest to the EV. The identification is done using the
    safety model as defined in \cref{s:safety_model} (line 9 of
    \cref{alg:attack}).
    
    \item Invokes the ``scenario matcher'' (SM) module (\circled{2} in
    \cref{fig:overview_ads_attack}), which uses the world state ($\mathbf{W}_t$)
    to determine whether the identified object is vulnerable to one of the attack
    vectors (as shown in \circled{3} and discussed in
    \cref{subsec:attack_vector}).

    \item Uses the ``safety hijacker'' (SH) (shown as \circled{4} in
    \cref{fig:overview_ads_attack}) to decide when to launch the attack ($t$),
    and for how long ($t+K$). The SH estimates the impact of the attack by using
    a shallow 3-hidden-layered NN, in terms of reduced safety potential
    ($\delta$).  The malware launches the attack {\it only} if the reduced
    safety potential drops below a predefined threshold (10 meters). We
    determine the threshold through simulation of driving scenarios that lead to emergency braking by the EV.  To evade detection, the malware ensures that $K$ does not exceed a
    pre-defined threshold (see line 15 in \cref{alg:attack}). $K$ is obtained by
    characterizing the continuous misdetection of an object associated with the
    ``object detector'' in the normal (i.e., without attacks) driving scenarios
    executed in the simulator.
\end{enumerate}

{\bf Phase 3. Triggering the attack.} \sysname:
\begin{enumerate}[noitemsep,nolistsep,leftmargin=*]
    \item   Uses the ``trajectory hijacker'' (\circled{5} in
    \cref{fig:overview_ads_attack}) to corrupt the camera feed. The trajectory
    hijacker perturbs the camera sensor data such that i) the trajectory of the
    object (e.g., a school bus) is altered to match the selected attack vector
    (e.g., Move\_Out), and ii) the trajectory of the object does not change
    significantly, thus evading detection.
    \item Attacks the trajectory of the victim object for the next $K$
    time-steps, chosen by the safety hijacker.
\end{enumerate}

\subsection{An Example of a Real Attack}
\cref{fig:attack-example} shows an example of a Move\_Out attack. Here we show
two different views: i) a simulation view, which was generated using a driving
scenario simulator, and ii) an ADS view, which was rendered using the
world-state visualizer.  

\sysname continuously monitors every camera frame using ``scenario matching''
(SM) to identify a target object for which the perceived trajectory by the EV
can be hijacked. If SM does not identify any target object of interest, it skips
the rest of the step and waits for the next camera frame. As shown in
\cref{fig:attack-example} (a) and (b), at time-step $t$, SM identified  an SUV
(i.e., target vehicle) as a target object of interest, and returned "Move\_Out"
as a matched attack vector, as the SUV was already in the Ego lane. Next,
\sysname launched "safety hijacker" to determine the reduced safety potential of
the attack and the number of time-steps the attack would need to be maintained.
As it turns out, the "safety hijacker" determined that the reduced safety
potential could cause an accident, so \sysname launched "trajectory hijacker" to
perturb the camera sensor data as shown in \cref{fig:attack-example} (c). Its
impact on the trajectory is shown in \cref{fig:attack-example}(d).  Camera
sensor data are perturbed by modifying individual pixels as shown in the white
area (in the bounding box (red square) of the target object) for illustration
purposes, because originally these pixels were modified in a way that was almost
invisible to the human eye. Because of this attack, the EV collided with the
target object as shown in \cref{fig:attack-example}(e) and (f).

 \begin{algorithm}[t!]
 \begin{algorithmic}[1]
 \renewcommand{\algorithmicrequire}{\textbf{Input:}}
 \renewcommand{\algorithmicensure}{\textbf{Output:}}
 \Require $\mathbf{\hat{S}_{t-1:t-2}}$ \Comment{Past object states} \Require
 $\mathbf{I_t^1}$ \Comment{Camera feed}
 \renewcommand{\algorithmicrequire}{\textbf{Global:}}
 \Require $attack$ \Comment{Flag indicating if the attack is active} \Require
 $K$ \Comment{Number of continuous attacks} \Require $i$ \Comment{Index of the
 target object} \Ensure $\mathbf{I_t^1}'$ \Comment{Perturbed image with
 adversarial patch} \State $\alpha \gets \varnothing $ \State $\mathbf{O_t},
 \mathbf{\hat{S}_{t}} \gets Perception (I_t)$ \If{$attack = False$} \State  $i,
 \delta_{t} \gets SafetyModel(\mathbf{\hat{S}_{t}})$ \Comment{From
 \cref{def:potential}} \State $\vec{v}_{rel, t}^{i} \gets
 calcVelocity({\hat{s}}_{t:t-1}^i)$ \State $\vec{a}_{rel, t}^{i} \gets
 calcAcceleration({\hat{s}}_{t:t-2}^i)$ \State $ \alpha \gets
 ScenarioMatcher({\hat{s}_{t}^{i}}) $ \EndIf

 \If{$\alpha \neq \varnothing $} \State  $attack, K \gets
    SafetyHijacker(\vec{a}_{rel,t}^{i}, \vec{v}_{rel,t}^{i}, \vec{\delta}_{t},
    \alpha$) \EndIf

\If{$attack = True~\land~K > 0$ } \State $\mathbf{I_t^1}' \gets
    TrajectoryHijacker(i, \mathbf{I_t^1}, {o_t^i}, {\hat{s}_{t-1}^i}, \alpha)$
    \State $K \gets K - 1$ \If {$K = 0$} \State  $attack \gets False$ \EndIf
    \EndIf
  \end{algorithmic}
 \caption{\label{alg:attack}Attack procedure at each time-step.}
\end{algorithm}
 \vspace{-0.2cm}
 \section{Algorithms and Methodology}\label{s:methodology}
In this section, we outline the three key steps taken by the malware: 
\begin{enumerate*}
\item in the monitoring phase, selecting the candidate attack vector by using
the scenario matcher (\cref{s:scenario_matcher}); 
\item in the monitoring phase, deciding when to attack by using the safety
hijacker  (\cref{s:time_of_attack}); and 
\item in the trigger phase, perturbing camera sensor feeds using the trajectory
hijacker.
\end{enumerate*} 
These steps are described in \cref{alg:attack}.

\subsection{Scenario Matcher: Selecting the Target Trajectory}~\label{s:scenario_matcher}
The goal of the scenario matcher is to check whether the closest object to the
EV (referred to as the \emph{target object}) is vulnerable to any of the
candidate attack vectors (i.e., Move\_Out, Move\_In, and Disappear). This is a
critical step for the malware, as it wants to avoid launching 1) an attack if
there are no objects next to or in front of the EV; or 2) an attack when the
object is actually executing the would-be bogus driving maneuver (e.g., selecting attack vector $\alpha$ = Move\_Out when the target is moving out of
the Ego lane anyway). The scenario-matching algorithm is intentionally designed
as a rule-based system (whose rules are listed in \cref{tab:scen-map})
to minimize its execution time, and hence evade detection. 

Note that "Scenario Matcher" can interchangeably choose between the Move\_Out
and Disappear attack vectors. However, in our work, we found that Disappear,
which requires a large perturbation in trajectory, is better suited for
attacking the pedestrians because the attack window is small. In contrast, the
attack window for vehicles is large. Therefore, for vehicles, \sysname uses
Move\_Out. This is described in detail in \cref{s:results}.

\subsection{Safety Hijacker: Deciding When to Attack}\label{s:time_of_attack}

To cause a safety violation (i.e., a crash or emergency brake), the malware will
optimally attack the vehicle when the attack results in $\delta~\leq~4$ meters.
The malware incorporates that insight into the safety hijacker to choose the
start and stop times of the attack by executing the safety hijacker at every
time-step. The safety hijacker at time-step $t$ takes ($\vec{v}_{rel,t}^{i},
\vec{a}_{rel,t}^{i}$), $\delta_t$, and $\alpha$ as inputs. It outputs the attack
decision (i.e., attack or no-attack) and the number of time-steps $K$ for which
the attack must continue to be successful (line 16 in \cref{alg:attack}). 

\begin{table}[!t]
\small
\centering
\begin{threeparttable}
\begin{tabular}{l|ll}
\toprule
TO trajectory            & \textbf{TO in EV-lane}             & \textbf{TO not
in EV-lane}          \\ 
\midrule
\textbf{Moving In}   & ---                 & Move\_Out/Disappear  \\ 
\textbf{Keep} & Move\_Out/Disappear & Move\_In             \\ 
\textbf{Moving Out}    & Move\_In            & ---                \\
\bottomrule
\end{tabular}
\begin{tablenotes}
\item TO: Target object
\end{tablenotes}
\end{threeparttable}
\caption{\label{tab:scen-map}Scenario Matching Map}
\end{table}

Let us assume that the malware has access to an oracle function $f_\alpha$ for a
given attack vector $\alpha$ that predicts the future safety potential of the EV
when it is subjected to the attack type $\alpha$ for  $k$ continuous time-steps,

\begin{equation}
    \delta_{t+k} = \mathit{f_\alpha}(\vec{v}_{rel,t}^i, \vec{a}_{rel,t}^i, \delta_t, k).
\label{eq:predict_delta}
\end{equation}

Later in this section, we will describe a machine-learning formulation that
approximates $f_\alpha$ using a neural network, and describe how to integrate it
with the malware. The malware decides to attack {\it only} when the safety
potential $\delta_{t+k}$ is less than some threshold $\gamma$.  Ideally, the
malware should attack when $\gamma = 4$ (i.e., corresponding to the
the $\delta$ for the crash).

In order to evade detection and disguise the attack as noise, the installed
malware should choose the "optimal $k$," which we refer to as $K$ (i.e., the
minimal number of consecutive camera sensor frame perturbations), using the
information available at time-step $t$. The malware can use the oracle function
$f_\alpha(.)$ to decide on the optimal number of time-steps ($K$) for which the
attack should be active. The malware decides to attack {\it only} if $k \le
K_{max}$, where $K_{max}$ is the maximal number of time-steps during which a
corruption of measurements cannot be detected. This is formalized in
\cref{eq:predict_k}. 
\begin{equation}
	K = \underset{k}{argmin}~k\cdot\left(\mathbb{I}(\delta_{t+k} \leq \gamma) = 1 \right)
\label{eq:predict_k}
\end{equation}

Finally, the malware must take minimal time to arrive at the attack decision.
However, in the current formulation, calculating $K$ can be very costly, as it
is necessary to evaluate \cref{eq:predict_k} using $f_\alpha$ (which is an NN)
for all $k \leq K_{max}$. We accelerate the evaluation of $K$ by leveraging the
fact that for our scenarios (\cref{s:scenarios}), $f_\alpha$ is non-increasing
with increasing $k$ when $\vec{a}_{{rel}_t} \le 0$. Hence, we can do a binary
search between $k \in [0, K_{max}]$ to find $K$ in $O(\log{}K_{max})$ steps.

\textbf{Estimating $\mathit{f_\alpha}$ using an NN.}
 We approximate the oracle function $f_\alpha$ using a feed-forward NN. We use
 an NN to approximate $f_\alpha$ to model the uncertainty in the ADS due to use
 of non-deterministic algorithms. Hence, the malware uses a uniquely trained NN
 for each attack vector. The input to the NN is a vector $[\delta_{t},
 \vec{v}_{rel_t}, \vec{a}_{rel_t}, k]$. The model predicts $\delta_{t+k}$ after
 $k$ consecutive frames, given the input. Intuitively, the NN learns the
 behavior of the ADS (e.g., conditions for emergency braking) and kinematics
 under the chosen attack vector, and it infers the safety
 potential ${\delta}_{t+k}$ to the targeted object from the input. We train the
 NN using a cost function ($\mathcal{L}$) that minimizes the average $L2$
 distance between the predicted ${\delta}_{{t+k}}$ and the ground-truth
 ${\delta^{G}_{{t+k}}}$ for the training dataset $\mathcal{{D}}_{train}$. 

\begin{equation}
    \begin{aligned}
    \mathcal{L} =  \frac{1}{|\mathcal{{D}}_{train}|} \sum_{i \in \mathcal{{D}}_{train}} \|{\delta^{G,i}_{t+k}} - {\delta}^i_{{t+k}}\|^2_2 \\
    \end{aligned}
    \label{eq:nn_loss}
\end{equation}

We use a fully connected NN with 3 hidden layers ($100, 100, 50$ neurons), the
ReLU activation function, and dropout layers with a dropout rate of $0.1$ to
estimate $\mathit{f_\alpha}$. The specific architecture of the NN was chosen to
reduce the computational time for the inference with sufficient learning
capacity for high accuracy. The NN predicts the safety potential after the
attack within 1m and 5m for pedestrian and vehicles, respectively.

The NN was trained with a dataset $\mathcal{{D}}$ collected from a set of
driving simulations ran on Baidu's Apollo ADS. To collect training data, we ran
several simulations, where each simulation had a predefined $\delta_{inject}$
and a $k$, i.e., an attack started as soon as the $\delta_{t} =
\delta_{inject}$, and continued for $k$ consecutive time-steps. The dataset
characterized the ADS's responses to attacks. The network was trained using the
Adam optimizer with a 60\%-40\% split of the dataset between the training and
the validation.

\subsection{Hijacking  Trajectory: Perturbing Camera Sensor Feeds}\label{s:adv_patch}
In this section, we describe the mechanism through which the malware can perturb
the camera sensor feeds to successfully mount the attack (i.e., execute one of
the attack vectors) once it has decided to attack the EV. The malware achieves
that objective using a trajectory hijacker.  

The attack vectors used in this paper require that the malware perturb the
camera sensor data (by changing pixels) in such a way that the bounding box
(${\hat{s}_{t}^i}$) estimated by the multiple-object tracker (used in the
perception module) at time $t$ moves in a given direction (left or right)
by $\omega_{t}$. 

The objective of moving the bounding box ${\hat{s}_{t}^i}$ in a given direction (left or right) can be formulated as an optimization problem. To solve it, we
modify the model provided by Jia {et al.}\cite{jia2019fooling} to evade attack
detection.  We find the translation vector $\omega_t$ at time $t$ that
maximizes the cost  $\mathbf{M}$ of Hungarian matching (recall $\mathbf{M}$ from
\cref{fig:overview_ads_attack}) between the detected bounding box,
${o_{t}^{i}}$, and the existing tracker state $\hat{s}_{t-1}^{i}$ (i.e., pushing the ${o_{t}^{i}}$ away from $\hat{s}_{t-1}^{i}$) such that the following conditions hold:
\begin{itemize}[noitemsep,nolistsep,leftmargin=*]
	\item Threshold $\mathbf{M} \le \lambda$ ensures that ${o_{t}^{i}}$ must
	still be associated with its original tracker state ${\hat{s}_{t-1}^{i}}$,
	i.e., $\mathbf{M} \le \lambda$. $\lambda$ can be found experimentally for a
	given perception system and depends on Kalman parameters. This condition is
	relaxed when the selected attack $\alpha=\text{"Disappear."}$ 
	\item $\omega_t \in [\mu-\sigma, \mu+\sigma]$ is within the Kalman
	noise parameters ($\mu, \sigma$) of the selected candidate object. This
	condition ensures that the perturbation is within the noise.
	\item Threshold $IoU({o}_{t}^{i} + \omega_t, ~patch) \ge \gamma$ ensures
	that the adversarial patch $patch$ should intersect with the detected
	bounding box, ${o_t^i}$, to restrict the search space of the patch. This condition can
	be removed when the attacker has access to the ADS, and can directly perturb $o_t^i$.
\end{itemize}
\begin{equation}
\begin{aligned}
\underset{\omega_{t}}{max}~\mathbf{M}~({o_{t}^{i}} + \omega_t,&~{\hat{s}_{t-1}^{i}})\\
\textit{s.t.} ~\mathbf{M} &\le \lambda, \\ ~IoU({o}_{t}^{i} + \omega_t, ~patch) &\ge \gamma,  \\
\omega_t &\in [\mu-\sigma, \mu+\sigma]
\end{aligned}
\label{eq:findpos}
\end{equation}
Finally, the malware should stop maximizing the distance between the
${o_{t}^{i}}$ and ${\hat{s}_{t-1}^{i}}$ when the object tracker has moved
laterally by $\Omega$ (i.e., the difference between the observed lateral distance
and the estimated lateral distance) since the attack start time $t-K'$. 

In our experiments, we found $K'$ to be generally around 4--20 frames, whereas
$K$ (i.e., total attack time determined by safety hijacker) was found to be generally
{10--65} frames. Since $K'$ is small, the chances of detection are significantly
lower.

\textbf{Perturbing Camera Sensor Data.} 
 Here the goal of the perturbation is to shift the position of the object
 detected by the object detector (e.g., YOLOv3). To achieve that objective, we
 formulate the problem of generating perturbed camera sensor data using Eq. (2)
 in \cite{jia2019fooling}. We omit the details because of lack of space. 

\subsection{Implementation}
We implemented \sysname using Python and C++. Moreover, we used fault
injection-based methods to implement the attack steps of \sysname. The proposed
malware has a small footprint, i.e., less than 500 lines of Python/C++ code, and
4\% additional GPU utilization with negligible CPU utilization in comparison to
the autonomous driving stack. This makes it difficult to detect an attack using
methods that monitor the usage of system resources.  \section{Experimental Setup}\label{s:implementation}

\subsection{AI Platform}
In this work, we use Apollo~\cite{apollo} as an AI agent for driving the AV.
Apollo is built by Baidu and is openly available on GitHub~\cite{apolloGit}.
However, we use LGSVL's version of Apollo 5.0~\cite{lgsvApolloGit}, as it
provides features to support integration of the LGSVL simulator~\cite{lgsvl}
with Apollo. Apollo uses multiple sensors: a Global Positioning System (GPS),
Inertial measurement units (IMU), radar, LiDAR, and multiple cameras. Our setup
used two cameras (fitted at the top and front of the vehicle) and one LiDAR for
perception.

\subsection{Simulation Platform}
We used the LGSVL simulator~\cite{lgsvl} that uses Unity~\cite{unity}, a gaming
engine ~\cite{gregory2017game}, to simulate driving scenarios. Note that a
driving scenario is characterized by the number of actors (i.e., objects) in the
world, their initial trajectories (i.e., position, velocity, acceleration, and
heading), and their waypoints (i.e., their route from source to destination). In
our setup, LGSVL simulated the virtual environment and posted virtual sensor
data to the ADS for consumption. The sensors included a LiDAR, a front-mounted
main camera, a top-mounted telescope camera, IMU, and GPS.
The measurements for different sensors were posted at different
frequencies~\cite{apollosensorconfig}. In our experiments, the cameras produced data at 15 Hz (of size 1920x1080), GPS at 12.5 Hz, and LiDAR
is rotating at 10 Hz and producing $360^{\circ}$ measurements per rotation. At
the time of this writing, LGSVL does not provide integration of continental
radar for Apollo. In addition, LGSVL provides Python APIs for creating driving
scenarios, which we leveraged to develop the driving scenarios described next.  

\subsection{Driving Scenarios}\label{s:scenarios}
\begin{figure*}
    \centering
    \includegraphics[width=\textwidth]{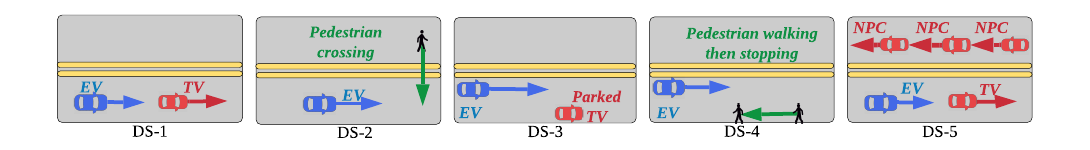}
    \caption{Driving scenarios. EV: Ego Vehicle. TV: Target Vehicle. NPC: Other Vehicles with no interaction with EV. }
    \label{fig:scenarios}
\end{figure*}

Here we describe the driving scenarios, shown in \cref{fig:scenarios}, that were
used in our experiments. All our driving scenarios were generated using LGSVL on
"Borregeas Avenue" (located in Sunnyvale, California, USA), which has a speed
limit of 50 kph. Unless otherwise specified, in all the cases EV was cruising at
45 kph. 

\textbf{In driving scenario 1} or "DS-1," the Ego vehicle (EV) followed a target
vehicle (TV) in the Ego lane at a constant speed (25 kph), as shown in Figure
\ref{fig:scenarios}. The TV started 60 meters ahead of the EV. In the golden
(i.e., non-attacked) run, the EV would accelerate to 40 kph and come closer to
the TV at the beginning, and then gradually decelerate to 25 kph to match the
speed of the TV. Thereafter, the EV maintained a longitudinal distance of 20 meters behind the TV for the rest of the scenario. We used this scenario to evaluate
the Disappear and Move\_Out attack vectors on a vehicle.

\textbf{In driving scenario 2} or "DS-2," a pedestrian illegally crossed the
street as shown in Figure \ref{fig:scenarios}.  In the golden run, the EV braked
to avoid collision and stopped more than 10 meters away from the pedestrian, if
possible. The EV started traveling again when the pedestrian moved off the road.
We used this scenario to evaluate the Disappear and Move\_Out attack vectors on
a pedestrian.

\textbf{In driving scenario 3} or "DS-3," a target vehicle was parked on the
side of the street in the parking lane. In the golden run, the EV maintained its
trajectory (lane keeping). We used this scenario to evaluate the Move\_In attack
vector on a vehicle.

\textbf{In driving scenario 4} or "DS-4," a pedestrian walked longitudinally
towards the EV in the parking lane (next to the EV lane) for 5 meters then stood
still for the rest of the scenario. In the golden run, EV recognized the
pedestrian, at which point it reduced its speed to 35 kph. However, once it
ensured that the pedestrian was safe (by evaluating its trajectory), it resumed
its original speed.  We use this scenario to evaluate the Move\_In attack vector on
a pedestrian.

\textbf{In driving scenario 5} or "DS-5," there are multiple vehicles with
random waypoints and trajectories, as shown in Figure \ref{fig:scenarios}.
Throughout the scenario, the EV was set to follow a target vehicle just as in
"DS-1," with multiple non-AV vehicles traveling on the other lane of the road as
well as in front or behind (not shown). Apart from the target vehicle, the
vehicles traveled at random speeds and starting from random positions in their
lanes. We used this scenario as the baseline scenario for a random attack to
evaluate the effectiveness of our attack end-to-end.

\subsection{Hardware Platform}
The production version of the Apollo ADS is supported on the
Nuvo-6108GC~\cite{nuvo6180}, which consists of Intel Xeon CPUs and NVIDIA GPUs.
In this work, we deployed Apollo on an x86 workstation with a Xeon CPU, ECC
memory, and two NVIDIA Titan Xp GPUs.
 \section{Evaluation \& Discussion}\label{s:results}

\subsection{Characterizing Perception System on Pretrained YOLOv3 in Simulation} \label{s:yolo_char}
\begin{figure*}
	\centering
    \captionsetup[subfigure]{oneside,margin={0.5cm,0cm}, justification=centering}
	\begin{subfigure}[b]{1.15in}
	    \centering
	    \includegraphics[width=1.15in, height=1.15in]{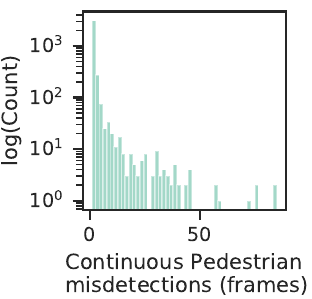}
	    \caption{Exp(loc = 1, $\lambda$ = 0.717) \\99th perc: 31.0}
	    \label{fig:yolov3baseline-a}
    \end{subfigure}
    \begin{subfigure}[b]{1.15in}
	    \centering
	    \includegraphics[width=1.15in, height= 1.15in]{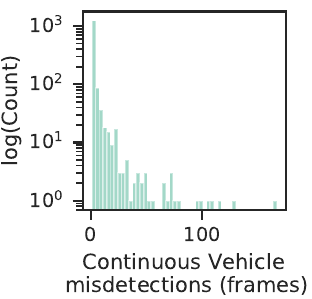}
	    \caption{Exp(loc = 1, $\lambda$ = 0.327) \\99th perc: 59.4}
	    \label{fig:yolov3baseline-b}
    \end{subfigure}
    \begin{subfigure}[b]{1.15in}
	    \centering
    	\includegraphics[width=1.15in, height= 1.15in]{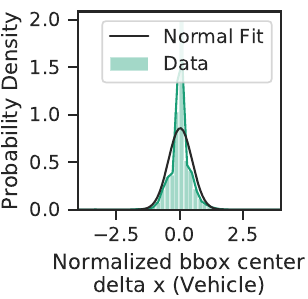}
	    \caption{Normal($\mu$ = 0.023, $\sigma$ = 0.464) \\99th perc: 1.145}
	    \label{fig:yolov3baseline-c}
    \end{subfigure}
    \begin{subfigure}[b]{1.15in}
	    \centering
	    \includegraphics[width=1.15in, height= 1.15in]{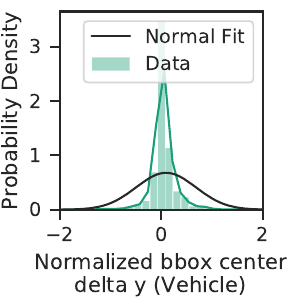}
	    \caption{Normal($\mu$ = 0.094, $\sigma$ = 0.586) \\99th perc: 1.775}
	    \label{fig:yolov3baseline-d}
    \end{subfigure}
    \begin{subfigure}[b]{1.15in}
	    \centering
    	\includegraphics[width=1.15in, height= 1.15in]{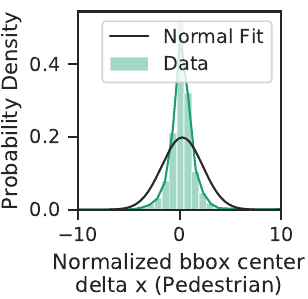}
	    \caption{Normal($\mu$ = 0.254, $\sigma$ = 2.010) \\99th perc: 5.235}
	    \label{fig:yolov3baseline-e}
    \end{subfigure}
    \begin{subfigure}[b]{1.15in}
	    \centering
    	\includegraphics[width=1.15in, height= 1.15in]{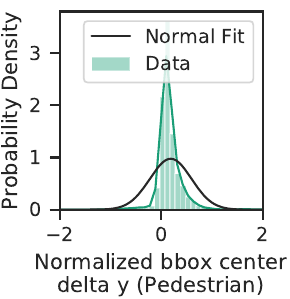}
	    \caption{Normal($\mu$ = 0.186, $\sigma$ = 0.409) \\99th perc: 1.868}
	    \label{fig:yolov3baseline-f}
    \end{subfigure}
    \hfill
    \vspace{-0.3cm}
	\caption{YOLOv3 object detection characterization on driving video generated using LGSVL. (a--b) show continuous misdetections with IoU=60\%. (c--f) show the distribution of normalized errors in the bounding box center predictions along the x and y coordinates of the image for vehicles and pedestrians.}
	\label{fig:yolov3baseline}
\end{figure*}

We characterize the performance of YOLOv3 (used in the Apollo perception system)
in detecting objects on the road, while the AV is driving, to measure
\begin{enumerate*}
\item the distribution of successive frames from an AV camera feed in which a
vehicle or a pedestrian is {\it continuously undetected}, and 
\item the distribution of {\it error in the center positions} of the predicted
bounding boxes compared to the ground-truth bounding boxes.
\end{enumerate*} 
We characterize those quantities to show that an attack mounted by \sysname and
the natural noise associated with the detector are from the same distribution.
In particular, we show that the continuous misdetection caused by \sysname is
within the 99th percentile of the continuous characterized misdetection
distribution of the YOLOv3 detector; see Figure \ref{fig:yolov3baseline}.  That
is important because if our attack fails, the object will reappear and be
flagged by the IDS as an attack attempt. Similarly, we characterize the error in
the predicted bounding box to ensure that our injected noise is within the
estimated Gaussian distribution parameters shown in Figure
\ref{fig:yolov3baseline}. \sysname changes the position at time-step $t$ by at
max $\mu -\sigma \le \omega \le \mu + \sigma$ of the Gaussian distribution.  For
this characterization, we generated a sequence of images and labels (consisting
of object bounding boxes and their classes) by manually driving the vehicle on
the San Francisco map for 10 minutes in simulation.

\par{\bf Continuous misdetections.}
\cref{fig:yolov3baseline} (a) and (b) show the distribution of the number of
frames in which pedestrians and vehicles were continuously misdetected. Here we
consider an object as misdetected if the IoU between the predicted and
ground-truth bounding boxes is less than $60\%$. The data follow an exponential
distribution.

\par{\bf Bounding box prediction error.}
To characterize the noise in the position of the bounding boxes predicted by
YOLOv3, we computed the difference between the center of the predicted bounding
box and the ground-truth bounding box and normalized it by the size
of the ground-truth bounding box. 

Only predicted bounding boxes that overlap with the ground-truth boxes are
considered. \cref{fig:yolov3baseline}(c), (d), (e), and (f) show the
distribution of normalized errors for the x (horizontal) and y (vertical)
coordinates in the image of the bounding box centers for pedestrians and
vehicles. The coordinates of the centers of the YOLOv3-predicted bounding boxes
follow a Gaussian noise model.  

\begin{table}[!t]
    \centering
\begin{threeparttable}
    \begin{tabular}{llllllll} 
    \toprule
         ID  & $K$ & \# runs & \# EB (\%) & \# crashes (\%)\\ 
    \midrule
         DS-1-Disappear-R & 48 & 101 & 54 (53.5\%) & 32 (31.7\%) \\ 
         DS-2-Disappear-R  & 14 & 144 & 136 (94.4\%) & 119 (82.6\%) \\ 
         DS-1-Move\_Out-R & 65 & 185 & 69 (37.3\%) & 32 (17.3\%) \\
         
         DS-2-Move\_Out-R  & 32  & 138 & 135 (97.8\%) & 116 (84.1\%)\\ 
         
         DS-3-Move\_In-R & 48 & 148 & 140 (94.6\%) & \textemdash \\ 
         
         DS-4-Move\_In-R & 24 & 135 & 106 (78.5\%) & \textemdash \\
         \textbf{DS-5-Baseline-Random}  & K* &  131 & 3 (2.3\%) & 0 \\
    \bottomrule
    \end{tabular}
    \label{tab:atk-res}
    \end{threeparttable}
	\caption{Smart malware attack summary compared with random (in bold). 
	EB: Emergency Braking. R: Robotack. 
	In our experiments, the AV tried emergency braking in all runs that resulted in accidents. 
	K* means K was randomly picked between 15 and 85 for each run of the experiment.} 
\end{table}
\subsection{Quantifying Baseline Attack Success}
In the baseline attack ({\it Baseline-Random}), we altered the object
	trajectory by
	\begin{enumerate*}[label=(\roman*)]
	\item randomly choosing an object (i.e., a vehicle or a pedestrian) for
	which the trajectory will be changed,
	\item randomly choosing the attack vector for a simulation run, 
	\item randomly initiating the attack at time-step $t$ of the driving scenario, and 
	\item continuing the attack for (a randomly chosen) $K$ time-steps.
	\end{enumerate*} 
	In other words, {\it Baseline-Random}) attack neither used scenario matcher nor used safety
	hijacker to mount the attack on the AV. However, it mimics trajectory
	hijacker to modify the trajectory of the vehicle.
	We used 131
	experimental runs of DS-5 in which the AV was randomly driving around the
	city to characterize the success of the baseline attack. Across all 131
	experimental runs (see "DS-5-Baseline-Random," \cref{tab:atk-res}), the AV
	performed emergency braking (EB) in {\it only} 3 runs (2.3\%) and
	crashed 0 times. 

	Here we also compare \sysname with attacks where both scenario matcher and trajectory hijacker
	are used (labeled as "R w/o SH" in
	\cref{fig:ds1234res}). However, these attacks do not use safety hijacker (SH). Hence, in
	these attacks, we randomly initiated the attack, and continue to
	attack for (a randomly chosen) $K$ time-steps, where $K$ is between 15 and
	85. The results of these attacks are described in detail in
	\cref{ss:safety_hijacker_impact}.

	Taken together these experiments provide a comparison with the current
	state-of-the-art adversarial attacks~\cite{jia2019fooling,lu2017adversarial}.

\subsection{Quantifying \sysname Attack Success}

In \cref{tab:atk-res}, \emph{ID} stands for the unique identifier for
experimental campaigns, which is a concatenation of "driving scenario id" and
"attack vector." Here a \emph{campaign} refers to a set of simulation runs
executed with the same driving scenario and attack vector. We also append TH and
SH to the ID to inform the reader that both trajectory-hijacking and
safety-hijacking were enabled in these attacks. Other fields are $K$ (median
number of continuous perturbations), \# runs (number of experimental runs), \#
EB (number of runs that led to AV emergency braking), and \# crashes (number of
runs that led to AV accidents). For each \textit{<driving scenario, attack vector>} pair,
we ran 150 to 200 experiments, depending on the total simulation time; however,
some of our experimental runs were invalid due to a crash of the simulator or
the ADS. Those experiments were discarded, and only valid experiments were used
for the calculations.

Across all scenarios and all attacks, we found that \sysname was significantly
more successful in creating safety hazards than random attacks were. \sysname
caused $\mathbf{33\times}$ more forced emergency brakings compared to random
attacks, i.e., \sysname caused forced emergency braking in \textbf{75.2\%} of
the runs (640 out of 851); in comparison, random attacks caused forced emergency
braking in \textbf{2.3\%} (3 out of 131 driving simulations). Similarly, random
attacks caused \textbf{0} accidents, whereas \sysname caused accidents in
\textbf{52.6\%} of the runs (299 out of 568, excluding Move\_In attacks). 
  
Across all our experiments, \sysname had a higher success rate in attacking
pedestrians (\textbf{84.1\%} of the runs that involved pedestrians) than in
attacking vehicles (\textbf{31.7\%} of the runs that involved vehicles).

{\bf Safety hazards with pedestrians.} We observed that \sysname was highly
effective in creating safety hazards in driving scenarios DS-2 and DS-4, which
involve pedestrians. Here we observe that in DS-2 with Move\_Out attacks, the EV
collided with the pedestrian in 84.1\% of the runs. Also, those attacks led to
EV emergency braking in 97.8\% of the runs. In DS-2 with Disappear attacks, the
EV collided with the pedestrian in 82.6\% of the runs and led to emergency
braking in 94.4\% of the runs. Finally, in DS-4 with Move\_In attacks, we did
not see any accidents with a pedestrian as there was no real pedestrian in the EV
lane; however, the Move\_In attacks led to emergency braking in 78.5\% of the
runs. Note that emergency braking can be life-threatening and injurious to
passengers of the EV, so it is a valid safety hazard. Interestingly, our malware
needed to modify only 14 camera frames for DS-2 with Disappear attacks and 24
frames for DS-4 with Move\_In attacks to achieve such a high success rate in
creating safety hazards. 

{\bf Safety hazards with vehicles.} We observed that \sysname was less
successful in creating hazards involving vehicles (DS-1 and DS-3) than in
creating hazards involving pedestrians. The reason is that LiDAR-based object
detection fails to register pedestrians at a higher longitudinal distance, while
recognizing vehicles at the same distance. Although the pedestrian is recognized
in the camera, the sensor fusion delays the object registration in the EV world
model because of disagreement between the LiDAR and camera detections. For the
same reason, \sysname needs to perturb significantly more camera frames
contiguously in the case of vehicles than in the case of pedestrians. However,
our injections were still within the bounds of the observed noise in object
detectors for vehicles. Overall, Move\_Out attacks in DS-1 caused emergency
braking and accidents in 37.3\% and 17.3\% of the runs, respectively, whereas
for the same driving scenario, Disappear attacks caused emergency braking and
accidents in 53.5\% and 31.7\% of the runs, respectively. \sysname was able to
cause emergency braking in 94.6\% of the runs by using Move\_In attacks in the
DS-3 driving scenario.

\subsection{Safety Hijacker \& Impact on Safety Potential}
\label{ss:safety_hijacker_impact}
\begin{figure*}
	\centering
	\includegraphics[width=1.7in]{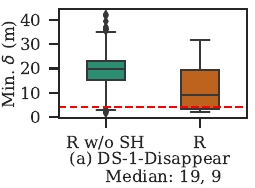}
	\includegraphics[width=1.7in]{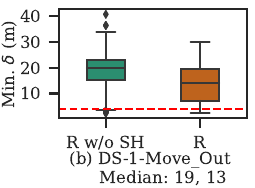}
	\includegraphics[width=1.7in]{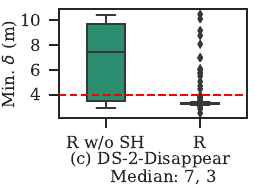}
	\includegraphics[width=1.7in]{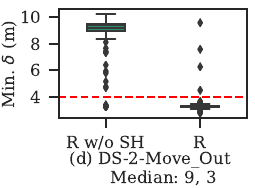}
	\caption{Impact of attacks. $\delta$: Safety potential. 
	`R': Robotack.
	`R w/o SH': Robotack without safety hijacker. 
	Dashed red line: Safety potential $\delta=4$ meters.}
	\label{fig:ds1234res}
\end{figure*}

Here we characterize the impact of attacks mounted by \sysname on the safety
potential of the EV with and without the safety hijacker (SH). Our results
indicate that the timing of the attack chosen by the SH is critical for causing
safety hazards with a high probability of success. In particular, with SH, the
number of successful attacks, i.e., forced emergency brakings and crashes, when
the vehicle trajectories are hijacked, were up to $\mathbf{5.1\times}$ and
$\mathbf{7.2\times}$ higher respectively, than the number of attacks induced at
random times using only trajectory hijacking. Attacks that hijacked pedestrian
trajectories were $\mathbf{14.8\times}$ and $\mathbf{24\times}$ more successful,
respectively.  \cref{fig:ds1234res} shows the boxplot of the minimum safety
potential of the EV measured from the start time of the attack to the end of the
driving scenario.  Recall that we label as an "accident" any driving scenario
that experiences a safety potential of less than 4 meters from the start of the
attack to the end of the attack. We determine the presence of forced emergency
braking by directly reading the values from the Apollo ADS. In
\cref{fig:ds1234res}, "R w/o SH" stands for "Robotack without SH", and "R" alone
stands for the proposed "Robotack" consisting of scenario matcher, trajectory
hijacker, and safety hijacker. Thus, the boxplot labeled "R w/o SH" indicates
that \sysname launched a trajectory-hijacking-attack on the EV without SH
(random timing), whereas "R" indicates that \sysname used the safety hijacker's
decided timing to launch a trajectory-hijacking-attack. We omit figures for the
Move\_In attack vector because in those scenarios the attacks did not reduce the
$\delta$ but caused emergency braking only. The improvements of "R" on attack
success-rate over "R w/o SH" are as follows.

\textbf{DS-1-Disappear.} \sysname caused  $7.2\times$ more crashes (31.7\% vs.
4.4\%). In addition, we observed that \sysname caused $4.6\times$ more emergency
braking (53.5\% vs. 11.6\%).

\textbf{DS-1-Move\_Out.} \sysname caused  $6.2\times$ more crashes (17.3\% vs.
2.8\%). In addition, we observed that \sysname caused $5.1\times$ more emergency
braking (37.3\% vs. 7.3\%).

\textbf{DS-2-Disappear.} \sysname caused  $7.9\times$ more crashes (82.6\% vs.
10.4\%). In addition, we observed that \sysname caused $2.4\times$ more
emergency braking (94.4\% vs. 39.4\%).

\textbf{DS-2-Move\_Out.} \sysname caused  $24\times$ more crashes (84.1\% vs.
3.5\%). In addition, we observed that \sysname caused $14.8\times$ more
emergency braking (97.8\% vs. 6.6\%).

\textbf{DS-3-Move\_In.} \sysname caused  $1.9\times$ more emergency brakings
(94.6\% vs. 50\%). A comparison of the number of crashes would not apply, as
there was no real obstacle to crash.

\textbf{DS-4-Move\_In.} \sysname caused  $1.6\times$ more emergency braking
(78.5\% vs 48.1\%). A comparison of the number of crashes would not apply, as
there was no real obstacle to crash.

{\textbf{Summary.}} In 1702 experiments (851 "R w/o SH", 851 "R") across all combinations of scenarios and attack vectors, \sysname (R) resulted in 640
EBs (75.2\%) over 851 "R" experiments. In comparison, \sysname without SH (R w/o SH) resulted in only
230 EBs (27.0\%) over 851 "R w/o SH" experiments. \sysname (R) resulted in 299
crashes (52.6\%) over 568 "R" experiments excluding DS-3 and DS-4 with Move\_In attacks, while \sysname without SH (R w/o SH) results in only 29 (5.1\%) crashes over the 568 "R w/o SH"
experiments excluding DS-3 and DS-4 with Move\_In attacks.

\begin{figure}[!t]
	\centering
	\includegraphics[width=1.7in]{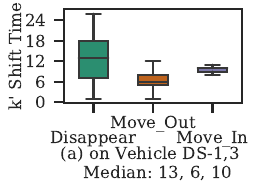}
	\includegraphics[width=1.7in]{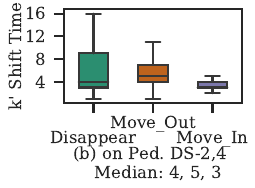}
	\caption{Time-steps $K'$ required to move object in/out by $\Omega$ (a) on vehicle, (b) on pedestrian.}
	\label{fig:kdash-nn-pred}
\end{figure}

\subsection{Evading Attack Detection}

Recall that the trajectory hijacker actively perturbs the estimated trajectory
for {\it only} $K' \ll K$ time-steps to shift the object position laterally at
most by $\Omega$, and it maintains the trajectory of the object for the next $K
- K'$ time-steps, where $K$ is the total number of time-steps for which the
attack must be active from start to end. Note that \sysname perturbs images for
all $K$ time-steps. However, \sysname modifies the image to change the
trajectory for only $K'$ time-steps, whereas for $K-K'$ time-steps, it maintains
the faked trajectory.

\cref{fig:kdash-nn-pred}(a) and (b) characterize $K'$ for different scenarios
and attack vectors. We observed that Move\_Out and Move\_In scenarios required a
smaller $K'$ in order to change the object position to the desired location than
the Disappear attack vector did. Furthermore, changing of a pedestrian's
location required smaller $K'$ than changing of a vehicle's location.

In those $K'$ time-steps, the disparity between between the Kalman filter's and
the object detector's output is not flagged as evidence of an attack because it
is within one standard deviation of the characterized mean during normal
situation.

\subsection{Characterizing Safety Hijacker Performance}
Here we characterize the performance of our neural network and its impact on the
malware's ability to cause a safety hazard. For lack of space, we discuss
results only for Move\_Out. 

\begin{figure}[!t]
    \centering
    \includegraphics[width=1.7in]{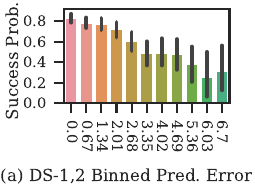}
    \includegraphics[width=1.7in]{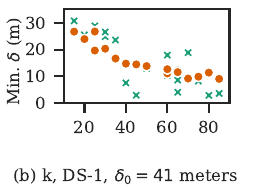}
    \caption{(a) NN binned prediction error for DS-1, DS-2 Move\_Out attack. (b) DS-1 Move\_Out attack, NN safety potential prediction. $\times$: ground-truth. $\bullet$: predicted. $k$: $\#$ of attacks. $\delta_0$: starting safety potential.}
    \label{fig:bin-err-pre} 
\end{figure}

\cref{fig:bin-err-pre}(b) shows a plot of the predicted value of the safety
potential (using the NN) and the ground-truth value of the safety potential
after the attack, as obtained from our experiments. We observe
that the predicted value is close to the ground-truth value of the safety
potential after the attack. On average, across all driving scenarios, NN's
prediction of the safety potential after the attack was within 5m and 1.5m of
the ground-truth values for vehicles and pedestrians, respectively.  

\cref{fig:bin-err-pre}(a) shows a plot of the probability of success (i.e., of
the malware's ability to cause a safety hazard) on the y-axis with increasing NN
prediction error probability on the x-axis. We find that the
success probability goes down as the prediction error of the safety potential
(using NN) increases. However, as stated earlier, our NN's prediction errors are generally
small.  \vspace{-0.2cm}
\section{Related Work}\label{s:related}
\vspace{-0.1cm}
\textbf{Security attacks.}
AVs are notoriously easy to hack into due to i) easy physical and software
access, ii) the large attack vectors available to the adversary because of the
complexity and heterogeneity of the software and hardware, and iii) the lack of
robust methods for attack detection. Hence, the insecurity of autonomous
vehicles poses one of the biggest threats to their safety and thus to their
successful deployment on the road.

\textbf{Gaining access to AVs.}
Hackers can gain access to the ADS by hacking existing software and hardware
vulnerabilities.  For example, research~\cite{winkelman2019autonomous,
koscher2010experimental} has shown that an adversary can gain access to the ADS
and launch cyber attacks by hacking vehicle-to-vehicle (V2V) and
vehicle-to-infrastructure (V2I) communication channels~\cite{sumra2011classes},
over-the-air software update mechanisms used by
manufacturers~\cite{sampath2007multi}, electronic component units
(ECUs)~\cite{koscher2010experimental}, infotainment
systems~\cite{InfotainmentHack}, and CAN buses~\cite{yaugdereli2015study}.
Another way of hacking an ADS is to use hardware-based malware, which can be
implanted during the supply chain of AVs or simply by gaining physical access to
the vehicle~\cite{koscher2010experimental}. In this work, we show an attack
approach that can masquerade as noise or faults and that can be implanted as
malware in either software or hardware.

\textbf{Adversarial machine learning and sensor attacks in AVs.}
A comparison with the current state-of-the-art adversarial ML-based attacks is provided
in \cref{s:intro}.

\textbf{Our attack.}
We find that none of the above work mentioned attacks geared toward i) evading
detection by an IDS or ii) explicitly targeting the safety of the vehicles. In
contrast, \sysname is the first attack approach that has been demonstrated on
production ADS software with multiple sensors (GPS, IMU, cameras, LiDAR) to
achieve both objectives by attacking only one sensor (the camera).

 \section{Conclusion \& Future Work}\label{s:conclusion}
\vspace{-0.05cm}
In this work, we present \sysname, smart malware that strategically attacks
autonomous vehicle perception systems to put the safety of people and property
at risk. The goal of our research is to highlight the need to develop more secure
AV systems by showing that the adversary can be highly efficient in targeting AV
safety (e.g., cause catastrophic accidents with high probability) by answering
the question of \textit{how, when and what to attack}. We believe that the broader
implications of our research are: 
\begin{enumerate*}[label=(\roman*)]
    \item Knowledge gathered from this
    type of study can be used to identify flaws in the current ADS architecture (e.g.,
    vulnerability in Kalman filters to adversarial noise) which can be used to drive
    future enhancements.
    \item Guide the development of countermeasures.
    \item Looking
    forward we believe these kinds of attacks can be fully automated and deployed as
    a new generation of malware.
\end{enumerate*}

The design of countermeasures is the subject of our future work. Existing
literature has shown a large number of adversarial attacks on these models
(e.g., object detection models and Kalman filters). Therefore, we are
investigating a broader solution that can dynamically and adaptively tune the
parameters of the perception system (i.e., parameters used in object detection,
Hungarian matching algorithm and Kalman filters) to reduce their sensitivity
to noise and thus, mitigate most of these adversarial attacks.

 \section*{Acknowledgments} \addcontentsline{toc}{section}{Acknowledgment}
\vspace{-0.1cm}
This material is based upon work supported by the National Science Foundation
(NSF) under Grant No. 15-35070 and CNS 18-16673. We thank our shepherd Kun Sun
for insightful discussion and suggestions. We also thank K. Atchley, J.
Applequist, Arjun Athreya, and Keywhan Chung for their insightful comments on
the early drafts. We would also like to thank NVIDIA Corporation for equipment
donation. Any opinions, findings, and conclusions or recommendations expressed
in this material are those of the authors and do not necessarily reflect the
views of the NSF and NVIDIA.
 \clearpage
{
    \balance
    \bibliographystyle{IEEEtran}
    \bibliography{references}
}

\end{document}